\documentclass{llncs}

\usepackage{amssymb}
\usepackage[english]{babel}
\usepackage[utf8x]{inputenc}
\usepackage{amsmath}
\usepackage{graphicx}
\usepackage{pxfonts}
\usepackage[colorinlistoftodos]{todonotes}
\usepackage{float}
\usepackage{epstopdf}
\usepackage{caption}
\usepackage{subcaption}
\usepackage{savesym}
\usepackage{llncsdoc}
\usepackage[colorlinks,linkcolor=black]{hyperref}
\usepackage{tabularx}
\usepackage{pgfplots}
\savesymbol{comment}
\restoresymbol{GNU}{comment}
\pgfplotsset{compat=1.14}

\newcommand{\newterm}[1]{{\emph{#1}}}
\newtheorem{asu}{Assumption}

\title{Safe Reinforcement Learning via Shielding}

\author{\small Mohammed Alshiekh\inst{1}, Roderick Bloem\inst{2}, R\"udiger Ehlers\inst{3}, Bettina K\"onighofer\inst{2}, Scott Niekum\inst{1}, Ufuk Topcu\inst{1}}

\institute{
   $^1$ University of Texas at Austin, USA \\
   $^2$ IAIK, Graz University of Technology, Austria\\
   $^3$ University of Bremen and DFKI GmbH, Bremen, Germany
          }

\usepackage{times}
\usepackage{booktabs}
\usepackage{rotating}
\usepackage{xcolor}
\definecolor{darkred}{rgb}{0.7, 0.0, 0.0}
\usepackage{caption}
\usepackage{listings}
\usepackage{xspace}
\usepackage{hyperref}
\usepackage{amsfonts}
\usepackage{amsmath}
\newcommand{\comment}[1]{}

\usepackage{amssymb}
\usepackage{temporal}
\usepackage{hyperref}
\usepackage{breakurl}
\usepackage{stmaryrd}
\usepackage{enumerate}
\usepackage{wrapfig} 

\newcounter{exacounter}

\usepackage{tikz}
\usetikzlibrary{arrows,automata,calc}
\tikzset{initial text={}}
\tikzset{every picture/.style=semithick} 
\tikzset{>=stealth'} 

\newcommand{\win}{\mathsf{win}}
\newcommand{\B}{\mathbb{B}}

\newcommand{\R}{\mathbb{R}}

\newcommand{\design}{\mathcal{S}}
\newcommand{\shield}{\mathcal{S}}
\newcommand{\model}{\mathcal{M}}
\newcommand{\pr}{\mathcal{P}}
\newcommand{\act}{\mathcal{A}}
\newcommand{\reward}{\mathcal{R}}
\newcommand{\states}{S}
\newcommand{\sstates}{Q}

\newcommand{\init}{q_0}
\newcommand{\din}{I}
\newcommand{\dinalph}{\Sigma_I}
\newcommand{\dinletter}{{\sigma_I}}
\newcommand{\dintrace}{{\overline{\sigma_I}}}
\newcommand{\dcombinedtrace}{{\overline{\sigma}}}
\newcommand{\dout}{O}

\newcommand{\doutalph}{\Sigma_O}
\newcommand{\doutletter}{{\sigma_O}}
\newcommand{\douttrace}{{\overline{\sigma_O}}}
\newcommand{\dalph}{\Sigma}
\newcommand{\dletter}{\sigma}
\newcommand{\dtrace}{\overline{\dletter}}
\newcommand{\lang}{\mathcal{L}}
\newcommand{\spec}{\varphi}
\newcommand{\gstates}{G}
\newcommand{\ginit}{g_0}
\newcommand{\game}{\mathcal{G}}

\newcommand{\LTLX}{\mathsf{X}}
\newcommand{\LTLG}{\mathsf{G}}
\newcommand{\LTLF}{\mathsf{F}}
\newcommand{\LTLU}{\mathsf{U}}
\newcommand{\NN}{\mathbb{N}}
\newcommand{\AP}{\mathsf{AP}}

\begin{document}
\maketitle
\begin{abstract}
\looseness-1
Reinforcement learning algorithms discover policies that maximize
reward, but do not necessarily guarantee safety during learning or execution pha\-ses.  We introduce a new approach to learn optimal policies while enforcing properties expressed in temporal logic.
To this end, given the temporal logic specification that is to be obeyed by the learning system, we propose to synthesize a reactive system  called a \emph{shield}.
The shield is introduced in the traditional learning process in two
alternative ways, depending on the location at which the shield is implemented.
In the first one, the shield acts each time the learning agent is about to make a decision and provides a list of safe actions.
In the second way, the shield is introduced after the learning agent. The shield monitors the
actions from the learner and corrects them only if the chosen action causes a violation of the specification.
We discuss which requirements a shield must meet to preserve the convergence guarantees of the learner.
Finally, we demonstrate the versatility of our approach on several challenging
reinforcement learning scenarios.
\end{abstract}

\section{Introduction}

Advances in learning have enabled a new paradigm for developing controllers for autonomous systems that are able to accomplish complicated tasks in possibly uncertain and dynamic environments. For example, in reinforcement learning (RL), an agent acts to optimize a long-term return that models the desired behavior for the agent and is revealed to it incrementally in a reward signal as it interacts with its environment \cite{SuttonB98}. Increasing use of learning-based controllers in physical systems in the proximity of humans also strengthens the concern of whether these systems will operate safely.

While convergence, optimality and data-efficiency of learning algorithms are relatively well understood, safety or more generally correctness during learning and execution of controllers has attracted significantly less attention. A number of different notions of safety were recently explored \cite{garcia15a,Pecka14}. We approach the problem of ensuring safety in reinforcement learning from a formal methods perspective. We begin with an unambiguous and rich set of specifications of what safety and more generally correctness mean. To this end, we adopt temporal logic as a specification
language \cite{Emerson:1991:TML:114891.114907}. For algorithmic purposes, we focus on the so-called {\it safety} fragment of (linear) temporal logic \cite{DBLP:books/daglib/0080029}. We then investigate the question ``how can we let, whenever it is fine, a learning agent do whatever it is doing, and also monitor and interfere with its operation whenever absolutely needed in order to ensure safety?''
In this paper, we introduce \newterm{shielded learning}, a framework that allows to apply machine learning to control systems in a way that the \emph{correctness} of the system's execution against a given specification is assured during the learning and controller execution phases, regardless of how fast the learning process converges.

\begin{figure}[tb]
\vspace{-18pt}
\begin{minipage}{\linewidth}
  \centering
  \begin{minipage}{0.44\linewidth}
    \begin{figure}[H]
      \centering
    \includegraphics[width=2.2in]{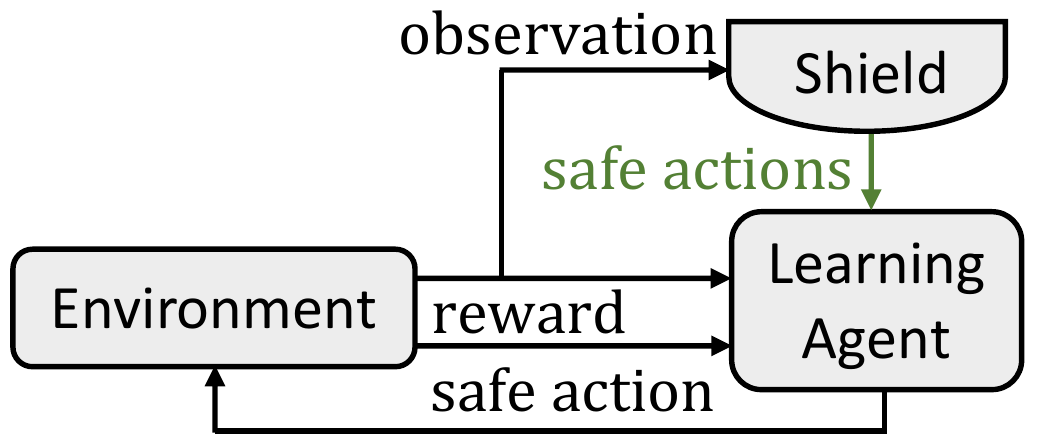}
    \caption{Preemptive Shielding.}
    \label{fig:shieldedlearner_before}
    \end{figure}
  \end{minipage}
  \hspace{0.03\linewidth}
  \begin{minipage}{0.45\linewidth}
    \begin{figure}[H]
      \centering
    \includegraphics[width=2.2in]{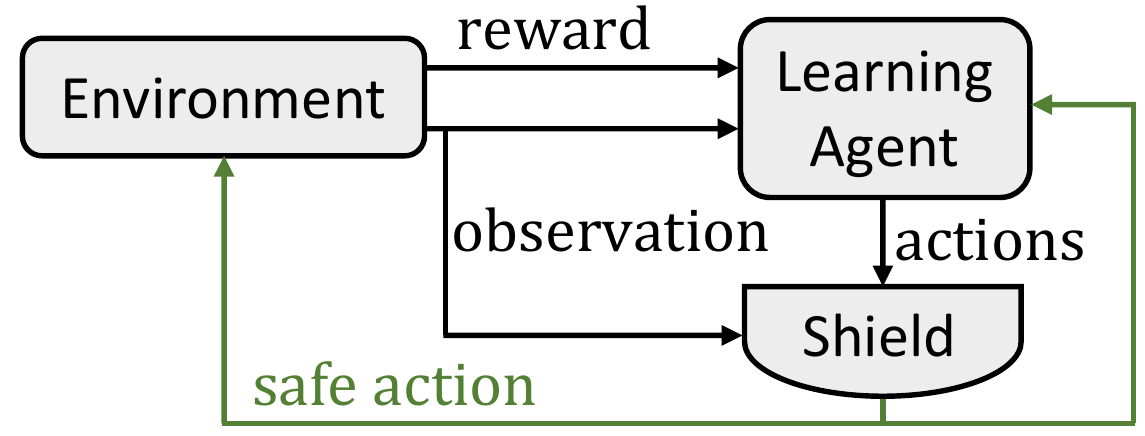}
    \caption{Post-Posed  Shielding.}
    \label{fig:shieldedlearner_after}
    \end{figure}
  \end{minipage}
\end{minipage}
\end{figure}

In the traditional reinforcement learning setting, in every time step, the learning agent chooses an action and sends it to the environment. The environment evolves according to the action and sends the agent an observation of its state and
a reward associated with the underlying transition. The objective of the learning agent is to optimize the reward accumulated over this evolution.

Our approach introduces a \newterm{shield} into the traditional reinforcement learning setting.
The shield is computed upfront from the safety part of the given system specification and an abstraction of the agent's environment dynamics. It ensures \emph{safety}
and \emph{minimum interference}. With minimum interference we mean that the shield restricts the agent as little as possible
and forbids actions only if they could endanger safe system behavior.

We modify the loop between the learning agent and its environment in two alternative ways, depending on the location at which the shield is implemented. In the first one, depicted in Fig.~\ref{fig:shieldedlearner_before}, the shield is implemented {\it before} the  learning agent and acts each time the learning agent is to make a decision and provides a list of {\it safe} actions. This list restricts the choices for the learner. The shield provides minimum interference, since it allows the agent to follow any policy as long as it is safe. In the alternative implementation of the shield, depicted in Fig.~\ref{fig:shieldedlearner_after},  it monitors the actions selected by the learning agent and corrects them if and only if the chosen action is unsafe.

Shielding offers several pragmatic advantages: Even though the inner working of learning algorithms is often complex, shielding with respect to critical safety specifications may be manageable (as we demonstrate in upcoming sections). The algorithms we present for the computation of shields make relatively mild assumptions on the input-output structure of the learning algorithm (rather than its inner working). Consequently, the correctness guarantees are agnostic---to an extent to be described precisely---to the learning algorithm of choice. Our setup introduces a clear boundary between the learning agent and the shield. This boundary helps to separate the concerns, e.g., safety and correctness on one side and convergence and optimality on the other and provides a basis for the convergence analysis of a shielded reinforcement learning algorithm.
Last but not least, the shielding framework is compatible with mechanisms such as
function approximation, employed by learning algorithms in order to improve their scalability.

\section{Related Work}

We now overview two complementing yet mostly isolated views on safety in reinforcement learning and in formal methods.

\paragraph{Safety in Reinforcement Learning.}
An exploration process is called \emph{safe} if no undesirable states
are ever visited, which can only be achieved through the incorporation of external knowledge \cite{garcia15a,moldovan2012safe}. The safety fragment of temporal logic that we consider is more general than the notion of safety of \cite{garcia15a} (which is technically a so-called \emph{invariance property} \cite{Baier:2008:PMC:1373322}).
One way of guiding exploration in learning is to provide \emph{teacher advice}.
A teacher (usually a human) provides advice (e.g., safe actions) when either the learner \cite{Pecka14,Clouse97} or the teacher \cite{Vidal13,Thomaz06} considers it to be necessary to prevent catastrophic situations.
For example, in a Q-learning setting, the agent
acts on the teacher's advice, whenever advice is provided. Otherwise, the agent chooses randomly between the set of actions with the highest Q-values. In each time step, the human teacher tunes the reward signal before sending it to the agent \cite{Thomaz06,ThomazB08}.
Our work is closely related to teacher-guided RL, since a shield can be considered as a teacher, who provides safe actions only if absolutely necessary.
In contrast to previous work, the reward signal does not have to be
manipulated by the shield, since the shield corrects unsafe actions in the
learning and deployment phases.

\paragraph{Safety in Formal Methods.}
Traditional correct-by-construction controller computation techniques are based on computing an abstraction of the environment dynamics and deriving a controller that guarantees
to satisfy the specification under the known environment dynamics.
Such methods combine \emph{reactive synthesis} with faithful environment modelling and abstraction. Wongpiromsarn et al.~\cite{DBLP:journals/tac/WongpiromsarnTM12} define a receding horizon control approach that combines continuous control with discrete correctness guarantees.
For simple system dynamics, the controller can be computed directly~\cite{DBLP:journals/tcs/HenzingerK99}. For more complex dynamics, both approaches are computationally too difficult. A mitigation strategy is to compute a set of low-level motion primitives to be combined to an overall strategy \cite{DBLP:conf/hybrid/DeCastroK16}. Having many motion primitives however also leads to inefficiency.
All of the above approaches have in common that a faithful, yet precise enough, abstraction of the physical environment is required, which is not only difficult to obtain in practice, but also introduces the mentioned computational burden. Control methods based on reinforcement learning partly address this problem, but do not typically incorporate any correctness guarantees. Wen et al.~\cite{DBLP:conf/iros/WenET15} propose a method to combine strict correctness guarantees with reinforcement learning.
They start with a non-deterministic correct-by-construction strategy and then perform reinforcement learning to limit it towards cost optimality without having to know the cost function a priori. Unlike the approach in the paper, their technique does not work with function approximation, which prevents it from being used in complex scenarios. Junges et al. \cite{DBLP:conf/tacas/Junges0DTK16} adopt a similar framework in a stochastic setting. A major difference between the works by Wen et al.~and Junges et al.~\cite{DBLP:conf/iros/WenET15,DBLP:conf/tacas/Junges0DTK16} on the one hand and the shielding framework on the other hand is the fact that the computational cost of the construction of the shield depends on the complexity of the specification and a very abstract version of the system, and is independent of the state space components of the system to be controlled that are irrelevant for enforcing the safety specification. Fu et al. \cite{DBLP:journals/tase/FuT16} establish connections between temporal-logic-constrained strategy synthesis in Markov decision processes and probably-approximately-correct-type bounds in learning \cite{DBLP:journals/cacm/Valiant84}.
Bloem et al. \cite{DBLP:conf/tacas/BloemKKW15} proposed the idea to synthesize a \emph{shield} that is attached to a system to enforce safety properties at run time.
We adopt this idea, and present our own realization of a shield, geared to the needs of the learning setting.

\section{Preliminaries}

We now introduce some basic concepts used in the following.

A \textbf{word} is defined to be a finite or infinite sequence of elements from some set $\Sigma$. The set of finite words over an alphabet $\Sigma$ is denoted by $\Sigma^*$, and the set of infinite words over $\Sigma$ is written as $\Sigma^\omega$. The union of $\Sigma^*$ and $\Sigma^\omega$ is denoted by the symbol $\Sigma^\infty$.

A \textbf{probability distribution} over a (finite) set $X$ is a function $\mu: X \rightarrow [0, 1] \subseteq \R$ with $\sum_{x\in X}\mu(x) = \mu(X) = 1$.
The set of all distributions on $X$ is denoted by $Distr(X)$.

A \textbf{Markov decision process} (MDP) $\model = (\states, s_I , \act, \pr, \allowbreak \reward)$ is a tuple
with a finite set $\states$ of states, a unique initial state $s_I \in \states$, a finite set $\act=\{a_1\dots a_n\}$ of Boolean actions,
a \emph{probabilistic transition function} $\pr: \states \times \act \rightarrow Distr (\states)$, and an
\emph{immediate reward function} $\reward: \states \times \act \times \states \rightarrow \R$.

In \textbf{reinforcement learning} (RL), an agent must learn a behavior through trial-and-error via interactions with an unknown environment modeled by a MDP $\model = (\states, s_I , \act, \pr, \reward)$.
Agent and environment interact in discrete time steps.
At each step $t$, the agent receives an observation $s_t$.
It then chooses an action $a_t\in\act$.
The environment then moves to a state $s_{t+1}$ with the probability $\pr(s_t, a_t, s_{t+1})$
and determines the reward $r_{t+1}=\reward(s_{t},a_{t},s_{t+1})$.
We refer to negative rewards $r_t<0$ as \emph{punishments}.
The \emph{return} $R = \sum_{t=0}^\infty \gamma^t r_t$ is the
cumulative future discounted reward, where $r_t$ is the immediate reward at time step $t$, and $\gamma \in [0, 1]$ is the \emph{discount factor} that controls the influence of future rewards.
The objective of the agent is to learn an \emph{optimal policy} $\Pi:\states\rightarrow\act$ that maximizes (over the class of policies considered by the learner) the expectation of the return; i.e. $max_{\pi\in\Pi}E_\pi(R)$, where $E_\pi(.)$ stands for the expectation w.r.t. the
policy $\pi$.

We consider a \textbf{reactive system} with a finite set
$\din=\{i_1,\ldots,i_m\}$ of Boolean input propositions and a finite set
$\dout=\{o_1,\ldots,o_n\}$ of Boolean output propositions.
The input alphabet is
$\dinalph=2^\din$, the output alphabet is $\doutalph=2^O$, and
$\dalph=\dinalph \times \doutalph$.
We refer to words over $\dalph$ as \emph{traces} $\dtrace$.  We write $|\dtrace|$ for the
length of a trace $\dtrace\in \dalph^{\infty}$.
A set of words $\lang
\subseteq \dalph^\infty$ is called a \emph{language}.

A \textbf{finite-state reactive system} is a tuple $\design = (\sstates, \init, \dinalph,
\doutalph, \delta, \lambda)$ with the input alphabet $\dinalph$, the output alphabet $\doutalph$, a
finite set of states $\sstates$, and the initial state $\init\in \sstates$. We assume that $\dinalph$ is a product of $\dinalph^1$ and $\dinalph^2$, i.e., we have $\dinalph =\dinalph^1 \times \dinalph^2$. Then, $\delta :
\sstates \times \dinalph \rightarrow \sstates$ is a complete transition
function, and $\lambda: \sstates \times \dinalph^1 \rightarrow \doutalph$
is a complete output function.  Given the input trace $\dintrace = (x^1_0,x^2_0)
(x^1_1,x^2_1) \ldots \in \dinalph^\infty$, the system $\design$ produces the
output trace $\douttrace = \design(\dintrace) = \lambda(q_0, x^1_0)
\lambda(q_1, x^1_1) \ldots \in \doutalph^\infty$, where $q_{i+1} =
\delta(q_i, (x^1_i,x^2_i))$ for all $i \ge 0$.
The input and output traces can be merged to the \emph{trace of $\mathcal{S}$} over the alphabet $\dinalph \times \doutalph$, which is defined as $\dcombinedtrace = ((x^1_0,x^2_0),\lambda(q_0, x^1_0)) ((x^1_1,x^2_1),\lambda(q_1, x^1)) \ldots \in (\dinalph \times \doutalph)^\omega$.

The finite-state reactive system definition is similar to that of a \emph{Mealy machine}, except that for choosing the output along a transition of the machine, only a part of the input is available. The larger generality of this model is needed for one type of shield that we introduce later, and such an extended Mealy-type computational model has already been used by Saqib and Somenzi~\cite{DBLP:journals/sttt/SohailS13} in the past.

\noindent
A \textbf{specification} $\spec$ defines a set $\lang(\spec) \subseteq
\dalph^\infty$ of allowed traces.
The reactive system $\design$ \emph{realizes} $\spec$, denoted by $\design \models \spec$, iff
$\lang(\design) \subseteq \lang(\spec)$.
$\spec$ is \emph{realizable} if there exists such an $\design$.
We assume $\spec$ is a set of
\emph{properties} $\{\spec_1,\ldots,\spec_l\}$ such that $\lang(\spec) =
\bigcap_i \lang(\spec_i)$. A system satisfies $\spec$ iff it
satisfies all its properties.

In most applications of formal methods, specifications of reactive systems are given as formulas in some \textbf{temporal logic}. \newterm{Linear temporal logic} \cite{DBLP:conf/focs/Pnueli77} (LTL) is a commonly used formal specification language. Given a set of propositions $\mathsf{AP}$, an LTL formula describes a language in $(2^\mathsf{AP})^\omega$. LTL extends Boolean logic by the introduction of temporal operators such as $\LTLX$ (next time), $\LTLG$ (globally/always), $\LTLF$ (eventually), and $\LTLU$ (until).
To use LTL for specifying a set of allowed traces by a reactive system, the joint alphabet $\Sigma = \Sigma_I \times \Sigma_O$ of the system must be decomposable into $\Sigma = 2^{\mathsf{AP}_I} \times \Sigma^\mathit{rest}_I \times 2^{\mathsf{AP}_O} \times \Sigma^\mathit{rest}_O$ for some system input and output components $\Sigma^\mathit{rest}_I$ and $\Sigma^\mathit{rest}_O$ that we do not want to reason about in the LTL specification. Then, the LTL formula can use $\mathsf{AP} = \mathsf{AP}_I \cup \mathsf{AP}_O$ as the set of atomic propositions. Given a trace $\dtrace$, we write $\dtrace_\AP$ to denote a copy of the trace where, in each character, the factors $\Sigma^\mathit{rest}_O$ and $\Sigma^\mathit{rest}_I$ have been stripped away so that $\dtrace_\AP \in (2^\AP)^\omega$.

Let us consider an example for an LTL specification that we build from ground up. By default, LTL formulas are evaluated at the first element of a trace. The LTL formula $r$ holds on a trace $\dtrace_\mathsf{AP} = \dtrace_0 \dtrace_1 \dtrace_2 \ldots \in (2^\AP)^\omega$ if and only if $r \in \dtrace_0$. The next-time operator $\LTLX$ allows to look one step into the future, so the LTL formula $\LTLX g$ holds if $g \in \dtrace_1$. We can take the disjunction between the formulas $r$ and $\LTLX g$ to obtain an LTL formula $(r \vee \LTLX g)$ which holds for a trace if at least one of $r$ or $\LTLX g$ hold. We can then wrap $(r \vee \LTLX g)$ into the temporal operator $\LTLG$ to obtain $\LTLG (r \vee \LTLX g)$. The effect of adding this operator is that in order for $\dtrace_\mathsf{AP}$ to satisfy $\LTLG (r \vee \LTLX g)$ is that $(r \vee \LTLX g)$ has to hold at every position in the trace. All in all, we can formalize this description by stating that we have that $\dtrace \models \LTLG (r \vee \LTLX g)$ holds if and only if for every $i \in \NN$, at least one of $r \in \dtrace_i$ and $g \in \dtrace_{i+1}$ hold.

A specification is called a \newterm{safety specification} if every trace $\dtrace$ that is not in the language represented by the specification has a prefix such that all words starting with the prefix are also not in the language.
Intuitively, a safety specification states that ``something bad should never happen''. Safety specifications can be simple \emph{invariance properties} (such as ``the level of a water tank should never fall below 1 liter''), but can also also be more complex (such as ``whenever a valve is opened, it stays open for at least three seconds'').
For specifications in LTL, it is known how to check if it is a safety language and how to compute a \newterm{safety automaton} that represents it \cite{DBLP:journals/fmsd/KupfermanV01}.

Such an automaton is defined as a tuple $\spec^s = (\sstates,
\init, \dalph, \delta, F)$, where $\dalph = \dinalph\times\doutalph$,
$\delta : \sstates \times \dalph \rightarrow \sstates$, and $F\subseteq
\sstates$ is a set of safe states.
 A \emph{run} induced by a trace
$\dtrace = \dletter_0 \dletter_1 \ldots \in \dalph^\infty$ is a sequence of
 states $\overline{q} = q_0 q_1 \ldots $ such that $q_{i+1} =
\delta(q_i, \dletter_i)$.  A trace $\dtrace$ of a system $\design$
\emph{satisfies} $\spec^s$ if the induced run visits only safe
states, i.e., $\forall i\geq 0 \scope q_i \in F$.

A (2-player, alternating) \textbf{game} is a tuple $\game = (\gstates,
\ginit, \dinalph, \doutalph, \delta, \win)$,
where $\gstates$ is a finite set of game states, $\ginit \in \gstates$ is the initial state,
$\delta: \gstates \times \dinalph \times \doutalph \rightarrow \gstates$
is a complete transition function, and $\win: \gstates^\omega
\rightarrow \B$ is a winning condition.  The game is played by the system and the environment.  In every state $g\in \gstates$
(starting with $\ginit$), the environment chooses an input
$\dinletter \in \dinalph$, and then the system chooses some output $\doutletter \in \doutalph$. These choices by the system and the environment define the next state $g' =
\delta(g,\dinletter, \doutletter)$, and so on. The resulting (infinite)
sequence $\overline{g} = g_0 g_1 \ldots$ is called a \emph{play}.  A play is \emph{won} by the system iff
$\win(\overline{g})$ is $\true$.
A (memoryless) \emph{strategy} for the system is a function $\rho:
\gstates \times \dinalph \rightarrow \doutalph$.
A strategy is
\emph{winning} for the system if all plays $\overline{g}$ that can be
constructed when defining the outputs using the strategy satisfy
$\win(\overline{g})$. The \emph{winning region} $W$ is the set of states
from which a winning strategy exists.

A \textbf{safety game} defines $\win$ via a set $F^g\subseteq \gstates$ of
safe states: $\win(g_0 g_1 \ldots)$ is $\true$ iff $\forall i \geq 0
\scope g_i \in F^g$, i.e., if only safe states are visited.
We will use safety games to
synthesize a \newterm{shield}, which implements the winning strategy in a new
reactive system $\shield = (\gstates, \init, \dinalph, \doutalph,
\delta', \rho)$ with $\delta'(g,\dinletter) =
\delta(g,\dinletter,\rho(g,\dinletter))$.

\section{Safety Specifications, Abstractions, and Game Solving}
\label{sec:SafetySpecs}

The goal of this paper is to combine the best of two worlds, namely
(1) the formal correctness guarantees of a controller with respect to a temporal logic specification, as provided by formal methods (and reactive synthesis in particular), and
(2) the optimality with respect to an a priori unknown performance criterion, as provided by reinforcement learning.

Consider the example of a path planner for autonomous vehicles.
Many general requirements on system behaviors such as safety
concerns may be known and expressed as specifications in temporal logic and can be enforced by reactive controllers.
This includes always driving in the correct lane, never jumping the red light, and never exceeding the speed limit~\cite{DBLP:conf/iros/WenET15}.
A learning algorithm is able to incorporate more subtle considerations, such as specific intentions for the current application scenario and
personal preferences of the human driver, such as reaching some goal
quickly but at the same time driving smoothly.

By combining reinforcement learning with reactive synthesis,
we achieve \emph{safe reinforcement learning}, which we define in
the following way:
\begin{definition}
Safe reinforcement learning is the process of learning an
optimal policy while satisfying a temporal logic safety specification $\spec^s$ during the learning and execution phases.
\end{definition}
In the following, we consider a safety specification to be given in the form of a deterministic safety
word automaton $\spec^s = (\sstates,
\init, \dalph, \delta, F)$, i.e., an automaton in which only safe states in $F$ may be visited. Note that since safety specifications given in linear temporal logic can be translated to such automata \cite{DBLP:journals/fmsd/KupfermanV01}, this assumption does not preclude the use of temporal logic as specification formalism.

Reactive synthesis enforces $\spec^s$ by solving a \emph{safety game} built from $\varphi^s$ and an abstraction of the environment in which the policy is to be executed. The game is played by the environment and the system.
In every state $q\in \sstates$, the environment chooses an input
$\dinletter \in \dinalph$, and then the system chooses some output $\doutletter \in \doutalph$. The play is won by the system if only safe states in $F$ are visited during the play.
In order to win, the system has to plan ahead:
it can never allow the play to visit a state from which the environment
can force the play to visit an unsafe state in the future.

Planning ahead is the true power of synthesis.
Let us revisit the autonomous driver example. Suppose that the car is
heading towards a cliff. In order to enforce that the car never crosses the cliff, it has to be slowed down long before it reaches the cliff, and thus far before an abnormal operating condition such as falling down can possibly be detected. In particular, the system has to avoid all states from which avoiding to reach the cliff
is no longer possible.

Planning ahead does not require
the environment dynamics to be completely known in advance.
However, to reason about when exactly a specification violation cannot be avoided, we have to give a (coarse finite-state) abstraction of the environment dynamics. Given that the environment is often represented as an MDP in reinforcement learning, such an abstraction has to be conservative with respect to the behavior of the real MDP. This approximation may have finitely many states even if the MDP has infinitely many states and/or is only approximately known.

Formally, given an MDP $\model = (\states, s_I , \act, \pr, \allowbreak \reward)$ and an \newterm{MDP observer function} $f : \states \rightarrow L$ for some set $L$, we call a deterministic safety word automaton $\spec^\model = (\sstates, \init, \dalph, \delta, F)$ an \newterm{abstraction} of $\mathcal{M}$ if $\Sigma = \act \times L$ and for every trace $s_0 s_1 s_2 \ldots \in \states^\omega$ with the corresponding action sequence $a_0 a_1 \ldots \in \act^\omega$ of the MDP, for every automaton run $\overline{q} = q_0 q_1 \ldots \in \sstates^\omega$ of $\spec^\model$ with $q_{i+1} = \delta(q_i,(l_i,a_i))$ for $l_i = L(s_i)$ and all $i \in \NN$, we have that $\overline{q}$ always stays in $F$.
An abstraction of an MDP describes how its executions can possibly evolve, and provides the needed information about the environment to allow planning ahead with respect to the safety properties of interest. Without loss of generality, we assume in the following that $\spec^\model$ has no states in $F$ from which all infinite paths eventually leave $F$. This requirement ensures that paths that model traces that cannot occur in $\mathcal{M}$ are rejected by $\spec^\model$ as early as possible.

The following example shows how specification automata and abstractions of MDPs are used.

\begin{figure}
\centering\begin{tikzpicture}

\draw[color=blue!80!white,fill=blue!70!white] (0,2.2) -- (0,0) -- (2,0) -- (2,2.2) ..controls +(-1,-0.3) and +(1,0.3) ..cycle;
\draw[color=black,line width=3pt] (0,2.5) -- (0,0) -- (2,0) -- (2,2.5) -- (1.5,2.75) -- (0.5,2.75) -- cycle;

\draw[color=black,line width=1.5pt] (0.85,2.75) -- (1.15,2.75) -- (1.15,3.0) -- (3.0,3.0) -- (3.0,3.3) -- (0.85,3.3) -- cycle;

\path[fill=white,line width=0pt] (0.85,2.55) -- (1.15,2.55) -- (1.15,3.0) -- (3.2,3.0) -- (3.2,3.3) -- (0.85,3.3) -- cycle;

\draw[line width=0.5pt,color=black!80!white] (2.7,2.98) -- (2.7,3.35) -- (2.55,3.50) -- (2.85,3.50) -- (2.7,3.35);

\draw[line width=0.5pt,color=red!80!black] (-0.2,0.25) -- (0.4,0.25) -- (0.15,0.35) -- (0.4,0.35) -- (0.15,0.45) -- (0.4,0.45) -- (0.15,0.55) -- (0.4,0.55) -- (0.15,0.65) -- (0.4,0.65) -- (0.15,0.75) -- (-0.2,0.75);

\draw[line width=1.5pt,color=black] (2.0,0.1) rectangle (3.0,0.4);
\path[line width=0,fill=blue!70!white] (1.8,0.1) rectangle ($(3.0,0.4)+(0.75pt,0)$);

\end{tikzpicture}
\caption{A hot water storage tank with an inflow, an outflow, and a tank heater.}
\label{fig:watertank}
\end{figure}
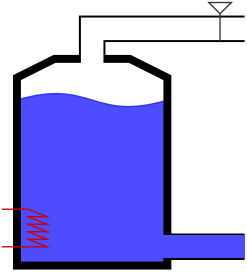

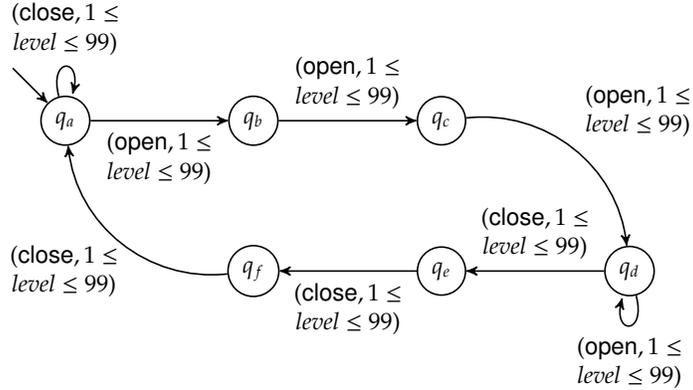
\begin{figure}
\centering\begin{tikzpicture}

\node[draw,shape=circle] (w1) at (0,0) {$q_a$};
\draw[->] (w1) edge[loop above] node[above] {\begin{tabular}{c}$(\mathsf{close},1 \leq$ \\ $\mathit{level} \leq 99)$ \end{tabular}} (w1);

\node[draw,shape=circle] (w2) at (2.5,0) {$q_b$};
\draw[->] (w1) edge node[below] {\begin{tabular}{c}$(\mathsf{open},1 \leq$ \\ $\mathit{level} \leq 99)$ \end{tabular}} (w2);

\node[draw,shape=circle] (w3) at (5,0) {$q_c$};
\draw[->] (w2) edge node[above] {\begin{tabular}{c}$(\mathsf{open},1 \leq$ \\ $\mathit{level} \leq 99)$ \end{tabular}} (w3);

\node[draw,shape=circle] (w4) at (7.5,-2) {$q_d$};
\draw[->] (w3) to[bend left=45] node[above right] {\begin{tabular}{c}$(\mathsf{open},1 \leq$ \\ $\mathit{level} \leq 99)$ \end{tabular}} (w4);

\draw[->] (w4) edge[loop below] node[below] {\begin{tabular}{c}$(\mathsf{open},1 \leq$ \\ $\mathit{level} \leq 99)$ \end{tabular}} (w4);

\node[draw,shape=circle] (w5) at (5.0,-2) {$q_e$};
\draw[->] (w4) edge node[above] {\begin{tabular}{c}$(\mathsf{close},1 \leq$ \\ $\mathit{level} \leq 99)$ \end{tabular}} (w5);

\node[draw,shape=circle] (w6) at (2.5,-2) {$q_f$};
\draw[->] (w5) edge node[below] {\begin{tabular}{c}$(\mathsf{close},1 \leq$ \\ $\mathit{level} \leq 99)$ \end{tabular}} (w6);

\draw[->] (w6) to[bend left=45] node[below left=-5pt] {\begin{tabular}{c}$(\mathsf{close},1 \leq$ \\ $\mathit{level} \leq 99)$ \end{tabular}} (w1);

\draw[->] (-0.7,0.7) -- (w1);

\end{tikzpicture}
\caption{The specification for the water tank controller. All states are accepting and transitions leading to the error state (which exists in addition to the states in the figure and is not accepting) are not shown.}
\label{fig:watertankSpec}
\end{figure}

\begin{figure}
\centering\begin{tikzpicture}

\node[shape=circle,draw] (q0) at (0,0) {$q_0$};
\node[shape=circle,draw] (q1) at (0,-3) {$q_1$};
\node[shape=circle,draw] (q2) at (0,-6) {$q_2$};
\node (q3) at (0,-9) {$\ldots$};
\node[shape=circle,draw] (q4) at (0,-12) {$q_{99}$};

\draw[->] (q0) edge[loop right] node[right] {\begin{tabular}{c}$(*,0 \leq$ \\ $\mathit{level} < 1)$ \end{tabular}} (q0);
\draw[->] (q1) edge[loop right] node[right] {\begin{tabular}{c}$(*,1 \leq$ \\ $\mathit{level} < 2)$ \end{tabular}} (q0);
\draw[->] (q2) edge[loop right] node[right] {\begin{tabular}{c}$(*,2 \leq$ \\ $\mathit{level} < 3)$ \end{tabular}} (q2);
\draw[->] (q4) edge[loop right] node[right] {\begin{tabular}{c}$(*,99 \leq$ \\ $\mathit{level} < 100)$ \end{tabular}} (q4);

\draw[->] (q0) to[bend left=7] node[right] {\begin{tabular}{c}$(\mathsf{open},1$ \\ $\leq \mathit{level} < 2)$ \end{tabular}}(q1);
\draw[->] (q1) to[bend left=7] node[left] {\begin{tabular}{c}$(\mathsf{close},0$ \\ $\leq \mathit{level} < 1)$ \end{tabular}}(q0);

\draw[->] (q1) to[bend left=7] node[right] {\begin{tabular}{c}$(\mathsf{open},2$ \\ $\leq \mathit{level} < 3)$ \end{tabular}}(q2);
\draw[->] (q2) to[bend left=7] node[left] {\begin{tabular}{c}$(\mathsf{close},1$ \\ $\leq \mathit{level} < 2)$ \end{tabular}}(q1);

\draw[->] (q2) to[bend left=7] node[right] {$\ldots$}(q3);
\draw[->] (q3) to[bend left=7] node[left] {$\ldots$}(q2);

\draw[->] (q3) to[bend left=7] node[right] {$\ldots$}(q4);
\draw[->] (q4) to[bend left=7] node[left] {$\ldots$}(q3);

\draw[->] (q0) ..controls +(3.5,-1.5) and +(3.5,1.5).. node[right] {\begin{tabular}{c}$(\mathsf{open},2$ \\ $\leq \mathit{level} < 3)$ \end{tabular}} (q2);

\node (skipper0) at (-2.5,-6) {$\ldots$};
\draw[->] (q1) ..controls +(-2.7,-1.5) and +(0,0.5).. node[above left=-2pt] {\begin{tabular}{c}$(\mathsf{open},3$ \\ $\leq \mathit{level} < 4)$ \end{tabular}}
(skipper0);

\node (skipper1) at (2.5,-9) {$\ldots$};
\draw[->] (q2) ..controls +(2.7,-1.5) and +(0,0.5).. node[above right=-2pt] {\begin{tabular}{c}$(\mathsf{open},4$ \\ $\leq \mathit{level} < 5)$ \end{tabular}}
(skipper1);

\node (skipper1) at (-2.5,-9) {$\ldots$};
\draw[->] (skipper1) ..controls +(0,-0.5) and +(-2.7,1.5).. node[below left=-2pt] {\begin{tabular}{c}$(\mathsf{open},99$ \\ $\leq \mathit{level} < 100)$ \end{tabular}}
(q4);
\end{tikzpicture}
\caption{The abstraction of the water tank behavior. All states are accepting and transitions leading to the error state (which exists in addition to the states in the figure and is not accepting) are not shown.}

\label{fig:watertankAbstraction}
\end{figure}
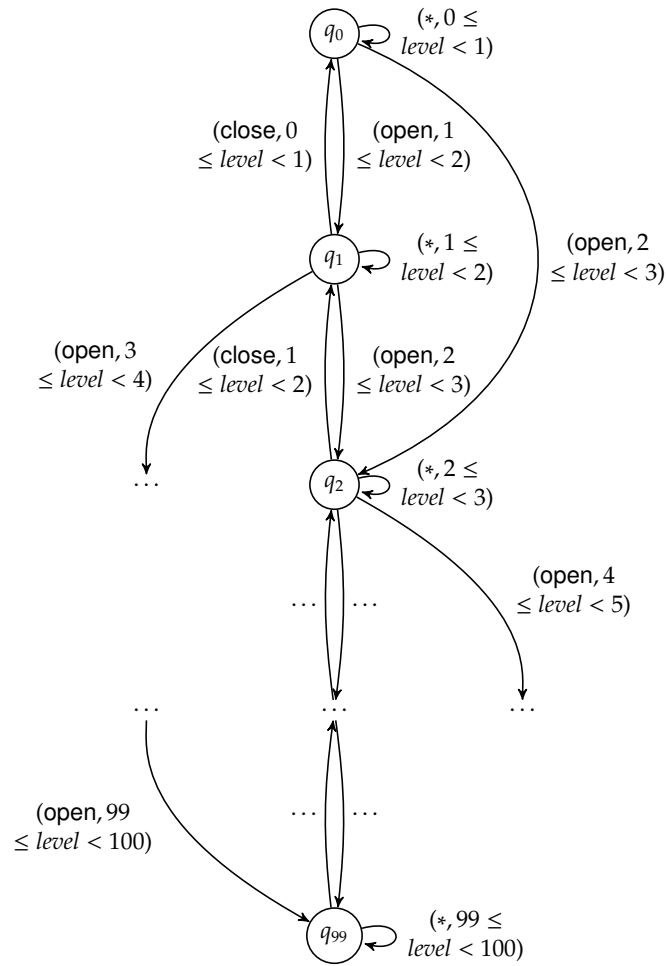

\begin{example}

We want to learn an energy-efficient controller for a hot water storage tank, depicted in Figure~\ref{fig:watertank}. Water stored in the task is kept warm by a heater whose energy consumption depends on the filling level of the tank, but we do not know what the exact relationship is.

The outflow is always between 0 and 1 liters per second, and the inflow is known to be between 1 and 2 liters per second whenever the valve is open (and it is 0 otherwise).
The capacity of the tank is limited to 100 liters, and whenever the inflow is switched on or off, the setting has to be kept for at least three seconds to limit the wear-out of the valve. Also, the tank must never overflow or run dry.

Let us now formalize this example. We can express the safety specification for the water tank valve controller using the following linear temporal logic formula:
\begin{align*}
 & \LTLG (\mathit{level}>0) \\
 \wedge \ & \LTLG(\mathit{level}<100) \\
 \wedge \ & \LTLG((\mathit{open} \wedge \LTLX \mathit{close}) \rightarrow \LTLX \LTLX \mathit{close} \wedge \LTLX \LTLX \LTLX\mathit{close}) \\
  \wedge \ & \LTLG((\mathit{close} \wedge \LTLX \mathit{open}) \rightarrow \LTLX \LTLX \mathit{open} \wedge \LTLX \LTLX \LTLX\mathit{open})
\end{align*}
The specification consists of four conjuncts, where the first two conjuncts enforce the water levels to be between the minimum and maximum thresholds. The next conjunct enforces that if the valve is open and then closed, then it has to stay closed for two more time steps (seconds). The final conjunct enforces that if the valve is closed and then opened, it has to stay open for two more time steps.

We can translate the specification to the safety automaton shown in Figure~\ref{fig:watertankSpec}. It uses the action sets $\act = \{\mathsf{open},\mathsf{closed}\}$ for the inflow valve state, and the label set $L = \{\mathit{level}< 1,1 \leq \mathit{level} \leq 99,\mathit{level}>99\}$ as needed information about the water tank filling status. What we know about the behavior of the water tank can be summarized as the abstraction automaton given in Figure~\ref{fig:watertankAbstraction}.

We will show in Section~\ref{sec:shield_synth} how to compute a shield from an abstraction automaton and a safety specification automaton. We will then revisit this example and give the resulting shield that enforces the specification. The shield will enforce that when the water level in the tank becomes too low, the inflow valve is opened until some minimum level of $4$ is reached, and it will also prevent the inflow from being opened when the level is above $93$. The latter is necessary as the valve has to stay open for at least three time steps. So as the inflow may be up to 2 liters/second during this time and the outflow may be 0, there is otherwise an overflow risk. As the shield is generated using the specification, it plans ahead for this not to happen, so it must prevent the opening of the inflow valve if the level is above $93$. Note that for more complicated specifications, the shield behavior can become much more complicated as well.
\end{example}

\section{Framework for Shielded Reinforcement Learning}
\label{sec:shieldedLearning}

In this section, we introduce a \emph{correct-by-construction} reactive system, called a shield, into the traditional learning process.
We propose two different ways to modify the loop between the learning agent and its environment: In Sec.~\ref{sec:preemptive}
we introduce the shield \emph{before} the learning agent.
In each time step, the shield modifies the list of actions available to the learner by providing a list of safe actions that the learning agent can choose from.
In Sec.~\ref{sec:postposed} the shield is implemented \emph{after} the learning agent. The shield monitors the actions selected by the
learning agent, and overwrites them if and only if the chosen action
is unsafe. Based on the location at which the shield is applied,
we call it \emph{preemptive} shielding and \emph{post-posed} shielding, respectively.
For both settings we make the following assumptions.
\begin{asu} \label{asu1}
    (i) The environment can be modeled as an MDP $\model = (\states, s_I , \act, \pr, \reward)$.
    (ii) We have constructed an abstraction $\spec^\model$.
    (iii) The learner accepts elements from $\states \times \sstates$ as state input (for the state space of the shield $\sstates$).
\end{asu}

We describe the operation of a learner and a shield together in this section, and give the construction for computing the shield in the next section. In both preemptive and post-posed shielding, the shield will be given as a reactive system $\design = (\sstates, \init, \dinalph,
\doutalph, \delta, \lambda)$.

\subsection{Preemptive Shielding}
\label{sec:preemptive}

\begin{figure}[tb]
\vspace{-18pt}
\begin{minipage}{\linewidth}
  \centering
  \begin{minipage}{0.43\linewidth}
    \begin{figure}[H]
      \centering
    \includegraphics[width=2.2in]{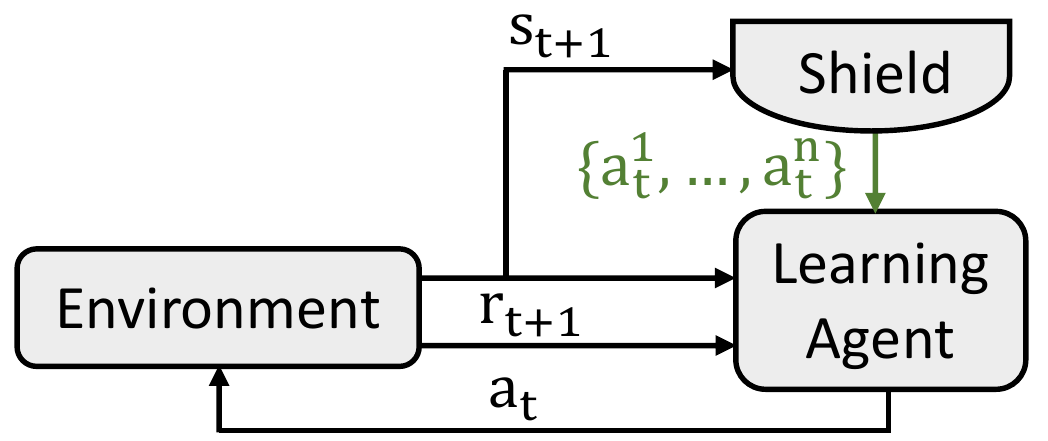}
    \caption{Preemptive Shielding.}
    \label{fig:shieldedlearner_before_detail}
    \end{figure}
  \end{minipage}
  \hspace{0.05\linewidth}
  \begin{minipage}{0.50\linewidth}
    \begin{figure}[H]
      \centering
    \includegraphics[width=2.4in]{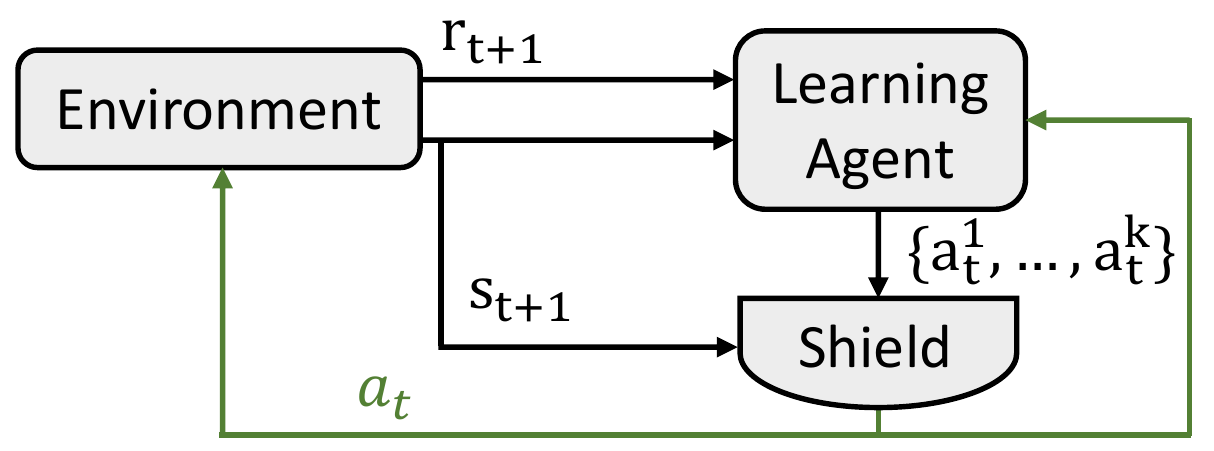}
    \caption{Post-Posed Shielding.}
    \label{fig:shieldedlearner_after_detail}
    \end{figure}
  \end{minipage}
\end{minipage}
\end{figure}

Fig.~\ref{fig:shieldedlearner_before_detail} depicts the preemptive
shielding setting.
The interaction between the agent, the environment and the shield is as follows:
At every time step $t$, the shield computes a set of all safe actions
$\{a_t^1,\dots,a_t^k\}$, i.e., it takes the set of all actions available, and removes all unsafe actions that would violate the safety specification $\spec_s$. The agent receives this list from the shield, and picks an action $a_t\in\{a_t^1,\dots,a_t^k\}$ from it. The environment executes action $a_t$, moves to a next state $s_{t+1}$, and provides the reward $r_{t+1}$.
The task of the shield is basically to modify the set of available
actions of the agent in every time step such that only safe actions remain.

More formally, for a preemptive shield, we have $\doutalph = 2^\mathcal{A}$, as the shield outputs the set of actions for the learner to choose from for the respective next step. The shield observes the label of the last MDP state in the sequence so far and provides the set of safe actions. For selecting the next transition of the finite-state machine that represents the shield, it also makes use of the action actually chosen by the agent. So for the input alphabet of the shield, we have $\dinalph = \dinalph^1 \times \dinalph^2$ with $\dinalph^1 = L$ and $\dinalph^2 = \mathcal{A}$.

The shield and the learner together produce a trace $s_0 a_0 s_1 a_1 \ldots \in (\states \times \act)^\omega$ in the MDP if there exists a trace $q_0 q_1 \ldots \in \sstates^\omega$ in the shield such that, for every $i \in \NN$, we have $a_i \in \lambda(q_i,L(s_i))$ and $q_{i+1} = \delta(q_i,(L(s_i),a_i)$.

\subsubsection{Properties of Preemptive Shielding.}
The preemptive shielding approach can also be seen as transforming the original MDP $\model$ into a new MDP $\model'=(\states', s_I, \act', \pr', \reward')$ with the unsafe actions at each state removed, and where $\states'$ is the product of the original MDP and the state space of the shield.
For each $s \in \states'$, we create a new subset of available actions $\act'_s\subseteq\act_s$ by applying the shield to $\act_s$ and eliminating all unsafe actions.
From each state $s \in \states'$, the transition function $\pr'$ contains only transition distributions from $\pr$ for actions contained in $\act'_s$.

\subsection{Post-Posed  Shielding}
\label{sec:postposed}

We propose a second shielding setting, in which the shield
is placed after the learning algorithm, as shown in Fig.~\ref{fig:shieldedlearner_after_detail}.
The shield monitors the actions of the agent,
and substitutes the selected actions by safe actions whenever this is necessary to prevent the violation of $\spec^s$.
In each step $t$, the agent selects an action $a_t^1$. The shield
forwards $a_t^1$ to the environment, i.e., $a_t = a_t^1$.
Only if $a_t^1$ is unsafe with respect to $\spec_s$,
the shield selects a different safe action $a_t \neq a_t^1$ instead.
The environment executes $a_t$, moves to $s_{t+1}$ and provides $r_{t+1}$.
The agent receives $a_t$ and $r_{t+1}$, and performs policy updates
based on that information. For the
executed action $a_t$, the agent updates its policy using $r_{t+1}$.  The question is what the reward for $a_t^1$ should
be in case we have $a_t \neq a_t^1$. We discuss two different approaches.
\begin{enumerate}
  \item \textbf{Assign a punishment $r'_{t+1}$ to $a_t^1$.}
    The agent assigns a punishment $r'_{t+1}<0$ to the unsafe action  $a_t^1$ and learns that selecting $a^1_t$ at state $s_t$ is unsafe, without ever violating $\spec^s$.
    However, there is no guarantee that unsafe actions are not part of the final policy.
    Therefore, the shield has to remain active even after the learning phase.

  \item \textbf{Assign the reward $r_{t+1}$ to $a_t^1$.}
    The agent updates the unsafe action $a^1_t$ with the reward $r_{t+1}$. Therefore, picking unsafe actions can likely be part of an optimal policy by the agent.
    Since an unsafe action is always mapped to a safe one, this does not pose a problem and the agent never has to learn to avoid unsafe actions.
    Consequently, the shield is (again) needed during the learning and execution phases.
\end{enumerate}

\subsubsection{Properties of Post-Posed Shielding.}

The big advantage of post-posed shielding is that it works even
if the learning algorithm is already in the execution phase and therefore follows a fixed policy.
In every step, the learning algorithm only sees the state of the MDP (without the state of the shield), and then the shield corrects the learner's actions whenever this is necessary to ensure safe operation of the system. The learning agent does not even need to know that it is shielded.

In order to be less restrictive to the learning algorithm, we propose that in every time step, the agent provides a ranking $rank_t=(a_t^1,\dots,a_t^k)$ on the allowed actions, i.e., the agent wants
$a^1_t$ to be executed the most, $a^2_t$
to be executed the second most, etc.
The ranking does not have to contain all available actions, i.e. $1\leq|rank_t|\leq n$, where
$n$ is the number of available actions in step $t$.
The shield selects the first action $a_t\in rank_t$
that is safe according to $\spec^s$. Only if all
actions in $rank_t$ are unsafe, the shield selects
a safe action $a_t \notin rank_t$.
Both approaches for updating the policy discussed before can naturally be extended for a ranking of several actions.
A second advantage of having a ranking on actions is that
the learning agent can perform several policy updates at once;
e.g., if all actions in $rank_t$ are unsafe, the agent can
perform $|rank_t|+1$ policy updates in one step by using the rewards $r'_{t+1}$ or $r_{t+1}$ for all of them, depending on which of the above variants is used.

\section{A Shield Synthesis Algorithm for Reinforcement Learning}
\label{sec:shield_synth}

A shield $\design$ is introduced into the traditional learning process, either before or after the learning agent.
In both cases, $\design$  enforces two properties: \emph{correctness} and \emph{minimum interference}.
First, $\design$ enforces correctness against a given safety specification $\spec^s$ at run time.
With minimum interference, we mean that the shield restricts the agent as rarely as possible.
The shield $\design$ is computed by reactive synthesis from $\spec^s$ and an MDP abstraction $\spec^\mathcal{M}$ that represents the environment in which the agent shall operate.

In this section, we give an algorithm to compute shields for preemptive shielding and post-posed shielding. We prove that the computed shields (1) enforce the correctness criterion, and (2) are the minimally interfering shields among those that enforce $\spec^s$ on all MDPs for which $\spec^\mathcal{M}$ is an abstraction.

The first steps of constructing the shield are the same for both variations of shielding.
Given is an RL problem in which an agent has to learn an optimal policy for an unknown environment that can be modelled by an MDP $\model = (\states, s_I , \act, \pr, \reward)$ while satisfying a safety specification $\spec^s = (\sstates, \init, \Sigma, \delta, F)$ with $\Sigma = \Sigma_I \times \Sigma_O$ and $\act=\doutalph$. We assume some abstraction $\spec^\model = (\sstates_\mathcal{M}, q_{0,\mathcal{M}}, \mathcal{A} \times L, \delta_\mathcal{M}, F_\mathcal{M})$ of $\model$ for some MDP observer function $f : S \rightarrow L$ to be given.
Since $\spec^s$ models a restriction of the traces of the MDP and the learner together that we want to enforce, we assume it to have $\Sigma = L \times \mathcal{A}$, i.e., it reads the part of the system behavior that the abstraction is concerned with.
We perform the following steps for both shield types.
\begin{enumerate}
  \item We translate $\spec^s$ and $\spec^\model$ to a safety game $\game = (\gstates, \ginit, \dinalph, \doutalph, \delta, F^g)$ between two players. In the game, the environment player chooses the next observations from the MDP state (i.e., elements from $L$), and the system chooses the next action. Formally, $\game$ has the following components:
  \begin{align*}
  G & = Q \times Q_\mathcal{M}, \\
  g_0 & = (q_0,q_{0,\mathcal{M}}), \\
  \Sigma_I & = L, \\
  \Sigma_O & = \mathcal{A}, \\
  \delta((q,q_\mathcal{M}),l,a) & = (\delta(q,(l,a)),\delta_\mathcal{M}(q,(l,a))), \\
   & \quad \text{for all } (q,q_\mathcal{M}) \in G, l \in L, a \in \mathcal{A}, \text{ and} \\
   F^g & = (F \times Q_\mathcal{M}) \cup (Q \times (Q_\mathcal{M} \setminus F_\mathcal{M})).
  \end{align*}
  In the construction, the state space of the game is the product between the specification automaton state set and the abstraction state set. The safe states in the game (in the set $F^g$) are the ones at which either the specification automaton is in a safe state, or the abstraction is in an unsafe state. The latter case represents that the observed MDP behavior differs from the behavior that was modeled in the abstraction. For game solving, it is important that such cases (whose occurrence in the field witnesses the incorrectness of the abstraction) count as winning for the system player, as the system player only needs to work correctly in environments that conform to the abstraction.
  \item Next, we compute the winning region $W\subseteq F^g$ of $G$ by standard safety game solving as described in \cite{DBLP:conf/tacas/BloemKKW15}.
\end{enumerate}
To compute a preemtive shield, we then perform the following step:
\begin{enumerate}
  \item[3.] We translate $G$ and $W$ to a reactive system $\design = (\sstates,_\design q_{0,\design}, \Sigma_{I,\design},
\Sigma_{O,\design}, \delta_\design, \lambda_\design)$ that constitutes the shield with $\Sigma_{I,\design} = \Sigma^1_{I,\design} \times \Sigma^2_{I,\design}$ for $\Sigma^1_{I,\design} = L$ and $\Sigma^2_{I,\design} = \mathcal{A}$. The shield has the following components:
  \begin{align*}
  \sstates_{\design} & = G, \\
  q_{0,\design} & = g_0, \\
  \Sigma_{I,\design} & = \mathcal{A} \times L, \\
  \Sigma_{O,\design} & = 2^\mathcal{A}, \\
  \delta_\design(g,l,a) & = \delta(g,l,a) \\
   & \quad \text{for all } g \in G, l \in L, a \in \mathcal{A}, \text{ and} \\
   \lambda_\design(g,l) & = \{ a \in \mathcal{A} \mid \delta(g,l,a) \in W\} \\
   & \quad \text{for all } g \in G, l \in L.
  \end{align*}
\end{enumerate}

\noindent To simplify $\design$, it makes sense to optionally remove all states that are unreachable from $q_{0,\design}$ after constructing $\design$.

\begin{figure}
\resizebox{\columnwidth}{!}{\begin{tikzpicture}

\node[draw,shape=circle] (c) at (-6,0) {$(q_3,q_d)$};
\node[draw,shape=circle] (b) at (-3,0) {$(q_2,q_e)$};
\node[draw,shape=circle] (a) at (0,0) {$(q_1,q_f)$};
\node[draw,shape=circle] (fail) at (3,0) {$q_\mathit{fail}$};

\node (aAlternative) at (0,-3) {$\ldots$};
\node (bAlternative) at (-3,-3) {$\ldots$};
\node (cAlternative) at (-9,-3) {$\ldots$};
\node (c2Alternative) at (-9,0) {$\ldots$};

\draw[->] (a) -- node[above] {$\mathit{open},\mathit{close}$} node[below] {$0 \leq \mathit{level} \leq 1$} (fail);
\draw[->] (b) -- node[above] {$\mathit{close}$} node[below] {$1 \leq \mathit{level} \leq 2$} (a);
\draw[->] (c) -- node[above] {$\mathit{close}$} node[below] {$2 \leq \mathit{level} \leq 3$} (b);
\draw[->] (b) to[bend left=30] node[above] {$\mathit{open}$}  (fail);
\draw[->] (a) to node[right] {\begin{tabular}{c}$\mathit{close}$, \\$1 \leq \mathit{level} \leq 2$ \end{tabular}} (aAlternative);
\draw[->] (b) to node[right] {\begin{tabular}{c}$\mathit{close}$, \\$2 \leq \mathit{level} \leq 3$ \end{tabular}} (bAlternative);
\draw[->] (c) to node[right] {\begin{tabular}{c}$\mathit{open}$, \\$4 \leq \mathit{level} \leq 5$ \end{tabular}} (cAlternative);
\draw[->] (c) to node[above] {\begin{tabular}{c}$\mathit{open}$, \\$5 \leq \mathit{level} \leq 6$ \end{tabular}} (c2Alternative);
\draw[->] (c) to[loop above] node[above] {\begin{tabular}{c}$\mathit{close}$, \\$3 \leq \mathit{level} \leq 4$ \end{tabular}} (c);

\end{tikzpicture}
} 
\caption{An excerpt for the product game of our running example. Transitions to paradise states are not shown.}
\label{fig:gameExcerpt}
\end{figure}
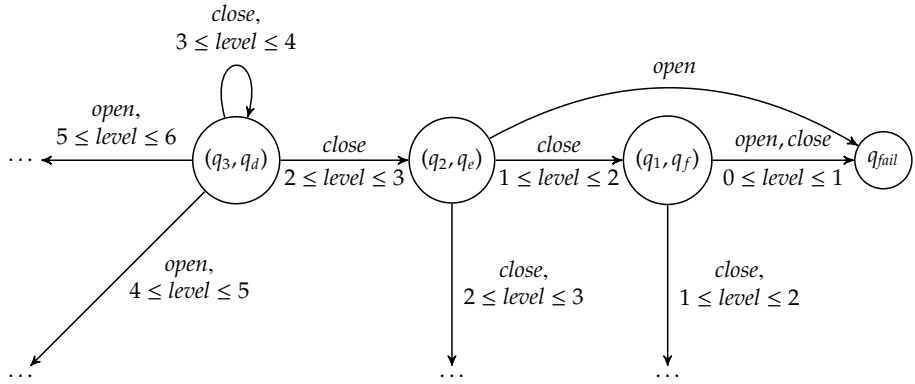

To exemplify these steps, let us reconsider the example from Section~\ref{sec:SafetySpecs}. Building the product game between the specification automaton and the MDP abstraction leads to a game with 602 states (if we merge all states in $F \times Q_\mathcal{M}$ into a single error state and all states in $Q \times (Q_\mathcal{M} \setminus F_\mathcal{M})$ into a single \hypertarget{paradisestatedef}{\textit{paradise state}} from which the game is always won by the system).
If we solve the game, then most of the states are winning, but a few are not.
Figure~\ref{fig:gameExcerpt} shows a small fraction of the game that contains such non-winning states. We can see that, in state $(q_3,q_d)$, the system should not choose action $\mathit{close}$, as otherwise the system cannot avoid to reach $q_\mathit{fail}$. It could be the case that $q_\mathit{fail}$ is actually not reached (when the environment chooses to let the level stay the same for a step), but we cannot be sure because we have to consider all evolutions of the environment to be possible that are consistent with our abstraction. Thus, the shield needs to deactivate the $\mathit{close}$ action in state $(q_3,q_d)$.

The shield allows all actions that are guaranteed to lead to a state in $W$, no matter what the next observation is. Since these states, by the definition of the set of winning states, are exactly the ones from which the system player can enforce not to ever visit a state not in $F$, the shield is minimally interfering. It disables all actions that may lead to an error state (according to the abstraction).

The construction of a post-posed shield is very similar to the construction of the preemptive shield. The main difference is that the post-posed shield always outputs a single action. Thus, the last step of the construction above should read as follows.

\begin{enumerate}
  \item[3.] We translate $G$ and $W$ to a reactive system $\design = (\sstates,_\design q_{0,\design}, \Sigma_{I,\design},
\Sigma_{O,\design}, \delta_\design, \lambda_\design)$ that constitutes the shield with $\Sigma_{I,\design} = \Sigma^1_{I,\design} \times \Sigma^2_{I,\design}$ for $\Sigma^1_{I,\design} = L \times \mathcal{A}$ and $\Sigma^2_{I,\design} = \{\cdot\}$. The shield has the following components:
  \begin{align*}
  \sstates_{\design} & = G, \\
  q_{0,\design} & = (q_0,q_{0,\mathcal{M}}), \\
  \Sigma_{I,\design} & = \mathcal{A} \times L, \\
  \Sigma_{O,\design} & = \mathcal{A}, \\
   \lambda_\design(g,l,a) & = \begin{cases}
   a & \text{if } \delta(g,l,a) \in W \\
   a' & \text{if } \delta(g,l,a) \notin W \text{ for some } \\
   & \text{ arbitrary but fixed } a' \text{ with } \delta(g,l,a') \in W,
  \end{cases} \\
  \delta_\design(g,l,a) & = \delta(g,l,\lambda_\design(g,l,a)) \\
  & \quad \text{for all } g \in G, l \in L, a \in \mathcal{A}. \\
  \end{align*}
\end{enumerate}
The construction can be extended naturally if a ranking of actions $\mathit{rank}_t=\{a^1_t,\dots,a^n_t\}$ is provided by
the agent. Then, the shield selects the first action $a_t = a^i_t$
that is allowed by $\spec^s$. Only if all actions in $\mathit{rank}_t$ are unsafe,
the shield is allowed to deviate and to select a safe action $a_t \notin rank_t$.

\subsection{Correctness and Minimal Interference of the Shields}

We now prove that the shields computed according to the definitions indeed have the claimed properties, namely {\it correctness}, and {\it minimal interference}. For brevity, we detail the case of preemptive shields. The line of reasoning for post-posed shielding is similar.

\paragraph{Correctness:}
A shield works correctly if for every trace  $s_0 a_0 s_1 a_1 \ldots \in (\states \times \act)^\omega$ that MDP, shield and learner can together produce, we have that $(f(s_0),a_0) (f(s_1),a_1) \ldots$ is in the language of the specification automaton $\varphi^S$ for the MDP labeling function $f$. Additionally, the shield must always report at least one available action at every step.

Let $q_0 q_1 \ldots \in \sstates^\omega$ be the run of $\varphi^S$ corresponding to $s_0 a_0 s_1 a_1 \ldots$, i.e., for which for every $i \in \NN$, we have $a_i \in \lambda(q_i,f(s_i))$ and $q_{i+1} = \delta(q_i,(f(s_i),a_i))$. By the construction of the shield, we have that $\sstates = Q \times Q_\mathcal{M}$, where $Q$ is the state space of $\varphi^S$ and $Q_\mathcal{M}$ is the state space of the abstraction. Hence, we can also write $q_0 q_1 \ldots$ as $(q^S_0,q^\mathcal{M}_0) (q^S_1,q^\mathcal{M}_1) \ldots$, where $q^\mathcal{M}_0 q^\mathcal{M}_1 \ldots$ is the run of the abstraction automaton on $s_0 a_0 s_1 a_1 \ldots$ (as defined in Section~\ref{sec:SafetySpecs}) and $q^S_0 q^S_1 \ldots$ is a run of $\varphi^S$ on $s_0 a_0 s_1 a_1 \ldots$. By the construction of the shield, it only has reachable states $(q^S,q^\mathcal{M})$ that are in the set of winning positions. For all possible next labels $l \in L$, there exists at least one action such that if the action is taken, then the next state $(q'^S,{q'}^\mathcal{M})$ is winning as well. Therefore, the shield cannot deadlock. As far as correctness is concerned, the $q^S$ component of the run of the shield will always reflect the state of the safety automaton along the trace, and since a winning strategy makes sure that only winning states are ever visited along a play, by the definition of $F^g$, the error state of $\varphi^S$ can only be visited after the error state for the abstraction MDP has been visited (and hence the abstraction turned out to be incorrect).

\paragraph{Minimal Interference:}
Let the shield, learner, and MDP together produce a prefix trace $s_0 a_0 s_1 a_1 \allowbreak{} s_2 \allowbreak{}a_2 \ldots s_k$ that induces a (prefix) run $q_0 q_1 \ldots q_{k-1} \in \sstates^*$ of the safety automaton $\varphi^S$ that we used as the representation of the specification for building the shield. Assume that the shield deactivates an action $a_{k+1}$ that is available from state $s_k$ in the MDP.
We show that the shield had to deactivate $a_{k+1}$ as there is another MDP that is consistent with the observed behavior and the abstraction for which, regardless of the learner's policy, there is a non-zero probability to violate the specification after the trace prefix $s_0 a_0 s_1 a_1 \allowbreak{} s_2 \allowbreak{}a_2 \ldots s_k a_{k+1}$.

Using the abstract finite-state machine $\varphi^\mathcal{M} = (Q_\mathcal{M},q_{0,\mathcal{M}},\Sigma,\delta,F)$, we define this other MDP $\mathcal{M}' = (S', s'_I, \mathcal{A}, \mathcal{P}', \mathcal{R})$ with $S' = Q_\mathcal{M} \times L$, $s'_I = q_{0,\mathcal{M}} \times f(s_0)$, $\mathcal{A}$ being the same set of actions as in the original MDP, and where $\mathcal{P}'((q,l),a) $ is a uniform distribution over all elements from the set $\{(q',l') \in Q_\mathcal{M} \times L \mid q' = \delta(q,(l,a)), q' \in F, \exists a' \in \mathcal{A}. \delta(q',(l',a')) \in F \}$ for every $(q,l) \in S'$ and $a \in \mathcal{A}$. Every state $(q',l') \in S'$ is mapped to $l'$ by the abstraction function $f$. The reward function is the same as in the original MDP, except that we ignore the (new) state component of the shield.

Assume now that action $a_{k+1}$ was activated after the prefix trace $s_0 a_0 s_1 a_1 \allowbreak{} s_2 \allowbreak{}a_2 \ldots s_k$ while the shield is in a state $(q^\mathcal{S}, q^\mathcal{M})$. We have that $\mathcal{M}'$ is an MDP in which every finite-length label sequence that is possible in the abstraction for some action sequence has a non-zero probability to occur if the action sequence is chosen. Due to the construction of the shield by game solving, action $a_{k+1}$ is only deactivated in state $(q^\mathcal{S}, q^\mathcal{M})$ if in the game, the environment player had a strategy to violate $\varphi^S$ using only traces allowed by the abstraction. Since  $\varphi^S$ is a safety property, the violation would occur in finite time. Since in $\mathcal{M}'$, all finite traces that can occur in the abstraction have a non-zero probability, activating $a_{k+1}$ (and the learner choosing $a_{k+1}$) would imply a non-zero proability to violate the specification in the future, no matter what the learner does in the future. Hence, the shield could not prevent a violation in such a case, and $a_{k+1}$ needs to be deactivated.

\section{Convergence}

Define an MDP $\model = (\states, s_I , \act, \pr, \reward)$, with discrete state set $S$, discrete state-dependent action sets $\act_s$, and state-dependent transition functions $\pr_s(a,s’)$ that define the probability of transitioning to state $s’$ when taking action $a$ in state $s$.
Assume also that a shield $\design = (\sstates_\design, q_{0,\design}, \Sigma_{I,\design},
\Sigma_{O,\design}, \delta_\design, \lambda_\design)$ is given for $\mathcal{M}$ and for some MDP labeling function $f : \states \rightarrow L$.

For both preemptive and post-posed shielding, we can build a product MDP $\mathcal{M'}$ that represents the behavior of the shield and the MDP together. Since $\mathcal{M}’$ is a standard MDP, all learning algorithms that converge on standard MDPs can be shown to converge in the presence of a shield under this construction.
Note that for the postposed shield case, this argument requires that whenever an action ranking is chosen by the learner that does not contain a safe action, there is a fixed probability distribution over the safe actions executed by the learner instead. This distribution may depend on the state of the MDP and the shield and the selected ranking, but must be constant over time, as otherwise we could not model the joint behavior of the shield and the environment MDP as a product MDP.

In both the post-posed and preemptive cases, we make use of the fact that the learner has access to the state of the shield and can base its actions on it in this argument. Shields can be relatively large---especially for complex abstractions and specifications---as they have both the state spaces of the abstraction and the specification automaton as factors. On the other hand, for specifications of the form ``at all points during the execution, the label of the MDP states should have a certain form'', the specification automaton has only a single state (plus an error state). The state space of the shield is then exactly the state space of the abstraction (plus \hyperlink{paradisestatedef}{paradise states} and error states). If the abstraction state can furthermore be determined from the respective last MDP state label, then the shield can be modified to have a single state (plus error states and paradise states). The requirements from Assumption~\ref{asu1} can then be relaxed by allowing the learner to only observe the state of the MDP (rather than the states of both the MDP and the shield) because, if the MDP behaves according to the abstraction, then the paradise state is never visited. At the same time, the shield ensures that no error state is ever visited. Hence, the state space of $\mathcal{M}'$ can be restructured to have to the same state space of $\mathcal{M}$.
In such a case, it suffices for the learner to observe the current state as state of $\mathcal{M}$ rather than $\mathcal{M}'$. To the learner, this is indistinguishable from operating on $\mathcal{M}$ without a shield.

\section{Experiments}

We applied shielded reinforcement learning in four domains: (1) a robot in 9x9 and 15x9 grid worlds, (2) a self-driving car scenario,  (3) an Atari\textsuperscript{\textregistered} game called \emph{Seaquest\texttrademark}, and (4) the water tank example from Section~\ref{sec:SafetySpecs}. For clarity, we compare between a subset of shielding settings which we later specify for each problem. The simulations were performed on a  computer equipped with an Intel\textsuperscript{\textregistered} Core\texttrademark  i7-4790K and 16 GB of RAM running a 64-bit version of Ubuntu\textsuperscript{\textregistered} 16.04 LTS. Source code, input files, and detailed instructions to reproduce our experiments are available for download.\footnote{\scriptsize\url{https://github.com/safe-rl/safe-rl-shielding}}

\subsection{Grid world Example}
We performed two experiments on a robot in a grid world. Snapshots of these environments are shown in Fig.~\ref{fig:grid-world}.
In both experiments, the robot's objective is to visit all the colored regions in a given order while maintaining one or both of the following safety properties.

\begin{itemize}
\item$\spec^s_1$: the robot must not crash into walls or the moving opponent agent. This specification applies to both experiments.
 \item$\spec^s_2$: the robot must not stay on a bomb for more than two consecutive steps. This specification applies only to the 9x9 experiment.
\end{itemize}
Fig.~\ref{fig:dfas} shows the deterministic finite automata corresponding to $\spec^s_1$ and $\spec^s_2$.

If the robot visits all marked regions in a given order (called episode), a reward is granted, and if a safety property is violated, a penalty is applied.
The agent uses tabular Q-learning with an $\epsilon$-greedy explorer that is capable of multiple policy updates at once.

\begin{figure}[!htb]
\vspace{-8pt}
\centering
    \begin{subfigure}[t]{0.26\linewidth}
    \centering
        \includegraphics[width=\linewidth,height=\linewidth]{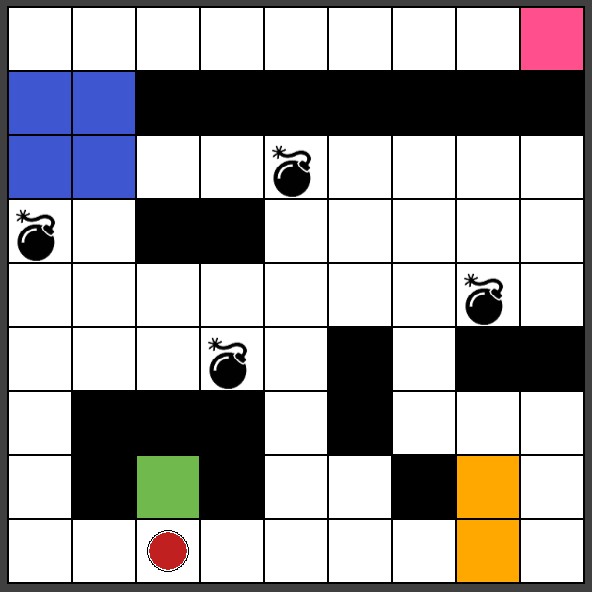}
        \label{fig:9x9 grid-worldC}
    \end{subfigure}
    \hspace{0.13\linewidth}
    \begin{subfigure}[t]{0.26\linewidth}
    \centering
        \includegraphics[width=\linewidth,height=\linewidth]{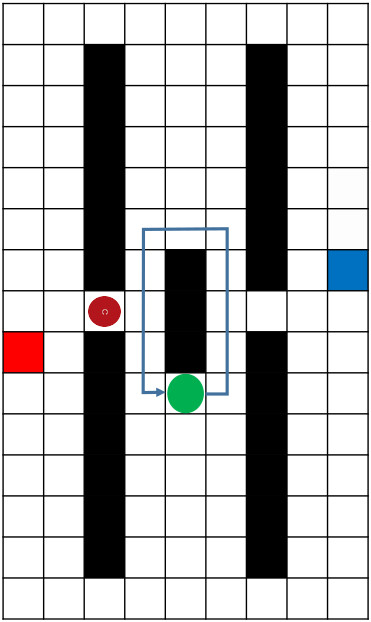}
        \label{fig:15x9 grid-world}
    \end{subfigure}
    \caption{Left: A 9x9 grid world with four bombs, a number of colored regions and walls (black) that must be visited in a specific order. Right: A 15x9 grid world with an opponent agent (green) circling the center of the grid-world and two colored regions that must be visited in a specific order. In both grid worlds, the robot (red) is only allowed to move north, south, east and west in each step.}
    \label{fig:grid-world}
   \end{figure}
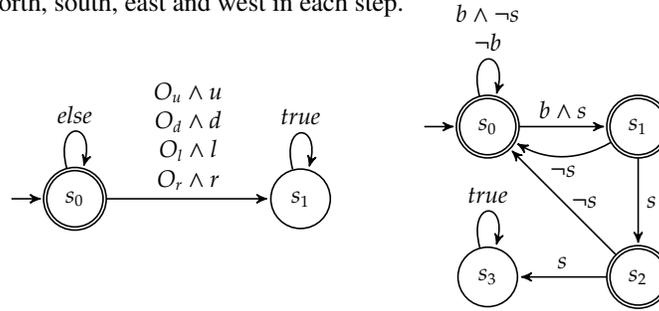
\begin{figure}
\vspace{-28pt}
\centering
\begin{subfigure}{0.4\linewidth}
\centering
\begin{tikzpicture}[>=stealth',shorten >=1pt,auto,node distance=3 cm, scale = 1, transform shape]

\node[initial,state,accepting] (A)                                    {$s_0$};
\node[state]         (B) [right of=A]                       {$s_1$};

\path[->]
      (A) edge [above]      node [align=center]  {$ O_u \wedge u $\\ $O_d \wedge d$\\ $O_l \wedge l$\\ $O_r \wedge r$} (B)
      (A) edge [loop above]       node [align=center]  {$ else $} (A)
      (B) edge [loop above]      node [align=center]  {$ true $} (B);

\end{tikzpicture}
\label{fig:dfa_wall}
\end{subfigure}
\begin{subfigure}{0.4\linewidth}
\centering
\begin{tikzpicture}[>=stealth',shorten >=1pt,auto,node distance=2 cm, scale = 1, transform shape]

\node[initial,state,accepting] (A)                                    {$s_0$};
\node[state,accepting]         (B) [right of=A]                       {$s_1$};
\node[state,accepting]         (C) [below of=B]                       {$s_2$};
\node[state]                   (D) [left of=C]                        {$s_3$};

\path[->]
      (A) edge [above]            node [align=center]  {$ b \wedge s$} (B)
      (A) edge [loop above]       node [align=center]  {$ b \wedge \neg s $ \\ $\neg b$} (A)
      (B) edge [bend left]        node [align=center]  {$ \neg s $} (A)
      (B) edge [right]            node [align=center]  {$ s $} (C)
      (C) edge [right]            node [align=center]  {$ \neg s $} (A)
      (C) edge [above]            node [align=center]  {$ s $} (D)
      (D) edge [loop above]       node [align=center]  {$ true $} (D);

\end{tikzpicture}

\label{fig:dfa_bomb}
\end{subfigure}
\caption{DFAs for $\spec^s_1$ (left) and $\spec^s_2$ (right).}
\label{fig:dfas}
\end{figure}

\pgfplotsset{every axis/.append style={thick},
label style={font=\Large},
tick label style={font=\large}  }

\begin{figure}[!htb]
\vspace{-4pt}
  \centering
  \begin{subfigure}{0.46\linewidth}
  \centering
\begin{tikzpicture}[scale=.65]
\centering
\begin{axis}[
xlabel={Episodes},
ylabel={Accumulated Reward},
grid=major,
legend pos=south east,
legend style={nodes={scale=0.85, transform shape}},
    y tick label style={
        /pgf/number format/.cd,
            fixed,
            fixed zerofill,
            precision=2,
        /tikz/.cd
    },
]
\addplot [red, dotted, very thick] table [
x=x,
y=no_shield,
] {data/9x9_illustrative.dat};
\addlegendentry{No shielding}
\addplot [gray, very thick] table [
x=x,
y=no_shield_neg,
] {data/9x9_illustrative.dat};
\addlegendentry{No shielding w/ Large penalty}
\addplot [green, very thick] table [
x=x,
y=shield_3_neg,
] {data/9x9_illustrative.dat};
\addlegendentry{ $|rank_t|=3$ w/ penalty}
\end{axis}
\end{tikzpicture}
  \end{subfigure}
  \hspace{0.03\linewidth}
  \begin{subfigure}{0.46\linewidth}
  \centering
\begin{tikzpicture}[scale=.65]
\centering
\begin{axis}[
xlabel={Episodes},
grid=major,
legend pos=south east,
legend style={nodes={scale=0.85, transform shape}},
    y tick label style={
        /pgf/number format/.cd,
            fixed,
            fixed zerofill,
            precision=2,
        /tikz/.cd
    },
]
\addplot [red] table [
x=x,
y=shield_1,
] {data/15x9_cycling.dat};
\addlegendentry{$|rank_t|=1$ w/o penalty}
\addplot [blue, dashed, very thick] table [
x=x,
y=shield_3,
] {data/15x9_cycling.dat};
\addlegendentry{$|rank_t|=3$ w/o penalty}
\addplot [red, dashed, very thick] table [
x=x,
y=shield_1_neg,
] {data/15x9_cycling.dat};
\addlegendentry{$|rank_t|=1$ w/ penalty}
\addplot [blue] table [
x=x,
y=shield_3_neg,
] {data/15x9_cycling.dat};
\addlegendentry{$|rank_t|=3$ w/ penalty}
\end{axis}
\end{tikzpicture}
  \end{subfigure}
\caption{The accumulated reward per episode for the 9x9 (left) and the 15x9 (right) grid worlds.}
\label{fig: grid-world plots}
\end{figure}
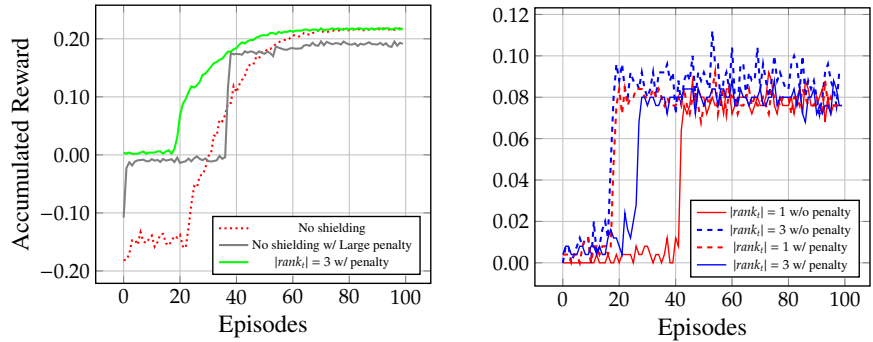

In the 9x9 grid-world, we synthesized a shield from $\spec^s_1 \wedge \spec^s_2$ and the (precise) environment abstraction in $2$ seconds. In the 15x9 experiment, we synthesized a shield from the (precise) environment abstraction and $\spec^s_1$ to prevent crashes into the wall and the moving opponent agent in $0.6$ seconds.

Fig.~\ref{fig: grid-world plots} shows that only the unshielded versions experience negative rewards. Furthermore, the shielded versions are not only safe, but also tend to learn more rapidly.
Whenever an unsafe action is picked, the agent updates at least two actions with a $|rank_t|=1$ shield, and up to 4 actions with a $|rank_t|=3$.
Fig.~\ref{fig: grid-world plots} (right) shows that only the shielded version $|rank_t|=3$ without penalty (blue, dashed) finds the optimal path, resulting in a higher average reward. In scenarios with $|rank_t|=1$ (red) or with penalties (solid), the agent computes a suboptimal path. In Fig.~\ref{fig: grid-world plots} (left), we compare between no shielding (red, dashed), no shielding with large penalties for unsafe actions (blue, solid), and a $|rank_t|=3$ post-posed shielding with penalties for corrected actions (green, solid). The unshielded version with large penalty does not reach the maximum reward score as the other two versions. In addition, the unshielded version does not speed up the learning of the agent as the  $|rank_t|=3$ does.

\subsection{A Self-Driving Car Example}

This example considers an agent that learns to drive around a block in a clockwise direction in an environment with the size of 480x480 pixels. In each step, the car moves 3 pixel in the direction of its heading and can make a maximum turn of $7.5$ degrees on the shortest direction to the commanded heading. After each step, the value of the reward and the new state of the car are returned. The state consists of the following four variables: the car's position in the x-axis, its position in the y-axis, the cosine and the sine of its heading. The safety specification in this example is to avoid crashing into a wall. The input to the shield is calculated from the car's state. It represents the side of the car with a distance less than 60 pixels away from any of the walls. Both of the preemptive and the post-posed shields were synthesized in $2$ seconds.
In each step, a positive reward is given if the car moves a step in a clockwise direction and a penalty is given if it moves in a counter-clockwise direction. A crash into the wall results in a penalty and a restart. The agent uses a Deep Q-Network (DQN) with a Boltzmann exploration policy. This network consists of four input nodes for the state variables, eight outputs nodes for the headings and three hidden layers.
\begin{figure}[!htb]
\centering
\vspace{-8pt}
\begin{minipage}{\linewidth}
\centering
\begin{subfigure}[t]{0.46\linewidth}
  \centering
\begin{tikzpicture}[scale=.65]
  \centering
\begin{axis}[
xlabel={Episodes},
ylabel={Accumulated Reward},
grid=major,
legend pos=south east,
legend style={nodes={scale=0.85, transform shape}},
]
\addplot [red, dotted,very thick] table [
	x=x,
	y=no_shield,
	] {data/car.dat};
  \addlegendentry{No shielding }
\addplot [blue] table [
	x=x,
	y=shield_1,
	] {data/car.dat};
 \addlegendentry{$|rank_t|=1$ w/o penalty}
\addplot [black, dotted, very thick] table [
	x=x,
	y=shield_1_pre,
	] {data/car.dat};
 \addlegendentry{Preemptive Shielding}
\end{axis}
\end{tikzpicture}
\label{fig:car plot}
\end{subfigure}
 \hspace{0.13\linewidth}
\begin{subfigure}[t]{0.38\linewidth}
\centering
\includegraphics[width=\linewidth, height=4.5cm]{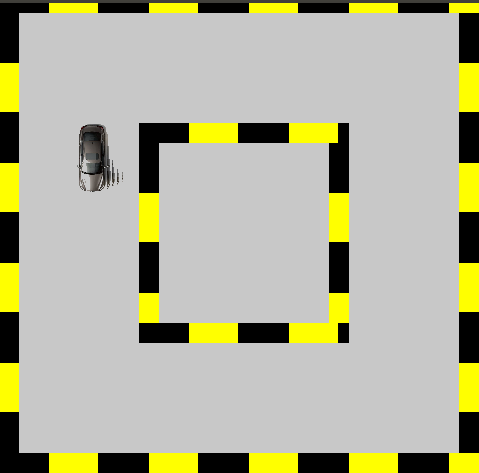}
\label{fig:car env}
\end{subfigure}
\end{minipage}
\caption{Left: The accumulated rewards per episode. Right: A snapshot of the environment, where the car is moving anti-clockwise.}
\label{fig:car}
\end{figure}
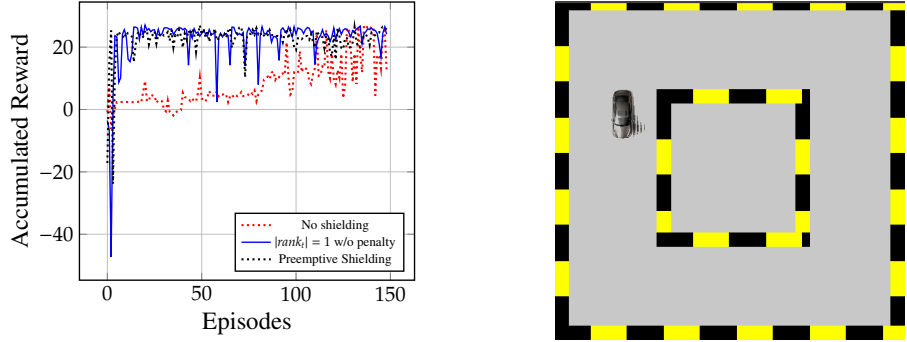

The plot in Fig.~\ref{fig:car}  shows that the accumulated rewards for unshielded reinforcement learning (red, dashed) increases over time, but still experiences crashes at the end of the simulation. The shielded version without punishment (blue, solid) learns more rapidly than the unshielded learning scenario and never crashes.

\subsection{Atari\textsuperscript{\textregistered} 2600 Seaquest\texttrademark}
\emph{Seaquest\texttrademark} is a underwater combat game in which the agent controls a submarine. The agent has to pick up divers under water, while avoiding or destroying
various objects, and must get to the surface before it runs out of oxygen. The goal of the agent is to maximize the game score.

\begin{figure}[!htb]
\vspace{-8pt}
\begin{minipage}{\linewidth}
  \centering
  \begin{subfigure}[t]{0.46\linewidth}
  \centering
\begin{tikzpicture}[scale=.65]
\centering
\begin{axis}[
xlabel={Episodes},
ylabel={Accumulated Reward},
grid=major,
legend pos=south east,
legend style={nodes={scale=0.85, transform shape}},
]
\addplot [red, dotted,very thick] table [
x=x,
y=no_shield,
] {data/seaquest.dat};
\addlegendentry{No shielding}
\addplot [blue] table [
x=x,
y=shield_1,
] {data/seaquest.dat};
\addlegendentry{ $|rank_t|=1$ w/o penalty}
\end{axis}
\end{tikzpicture}
\centering
\label{fig:seaquest plot}
\end{subfigure}
\hspace{0.15\linewidth}
\begin{subfigure}[t]{0.37\linewidth}
\centering
\includegraphics[width=\linewidth, height=4.8cm]{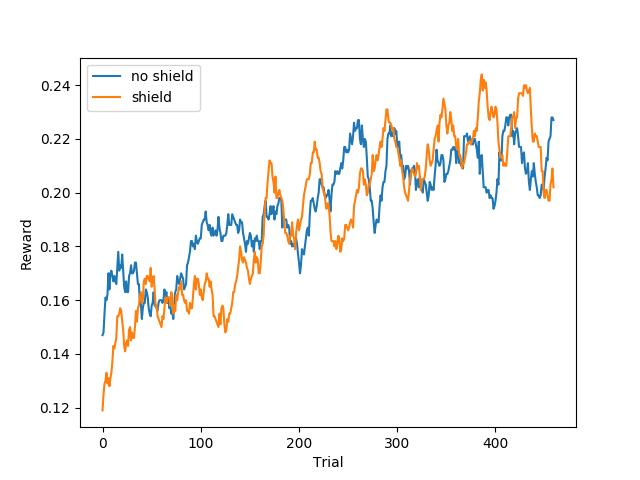}
\label{fig:seaquest env}
\end{subfigure}
\end{minipage}
\caption{Left: The accumulated rewards per episode. Right: A snapshot of \emph{Seaquest\texttrademark}.}
\label{fig:seaquest}
\end{figure}
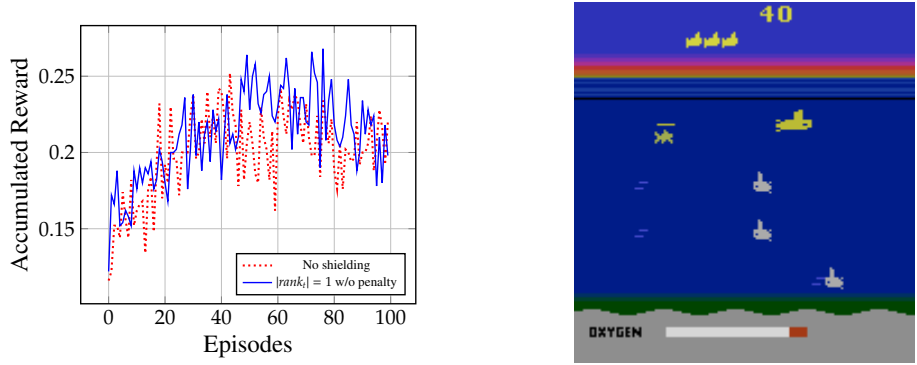
For our experiments, we used the OpenAI Gym\footnote[1]{https://gym.openai.com/}
library that integrates the Arcade Learning Environment (ALE)~\cite{13jair-ale}, and a Python implementation\footnote[2]{https://github.com/devsisters/DQN-tensorflow}
of DeepMind's Deep Reinforcement Learning approach~\cite{mnih-dqn-2015}.
The agent receives as input only RGB images of the screen as in Fig.~\ref{fig:seaquest} (right). The agent is used purely as a black box, only changing actions that violate the specification described below.

We model two simple safety properties. First,
the submarine has to surface before oxygen runs out ($\spec^s_1$). Secondly,
the submarine is not allowed to surface if it has enough oxygen but
has not collected any divers yet ($\spec^s_2$).
The specification $\spec^s=\spec^s_1\wedge\spec^s_2$ decides when the submarine
has to surface and when it is not allowed to surface, depending on the actual depth, the status of the oxygen reserves, and the number of collected divers.
We compute all inputs of the shield from the state of the Atari\textsuperscript{\textregistered} simulator.
The results illustrated in Fig.~\ref{fig:seaquest} (left) show that shielding the learner did not change its performance, however, the safety properties $\spec^s_1\wedge\spec^s_2$ were not violated when shielding the learner.

\subsection{The Water Tank Example}

In the example shown in Fig.~\ref{fig:watertank}, the tank must never run dry or overflow by controlling the inflow switch ($\spec^s_1$). In addition, the inflow switch must not change its mode of operation before 3 time steps have passed since the last mode change  ($\spec^s_2$). Refer to example 1 of section~\ref{sec:SafetySpecs}, for a full description of the abstract water tank dynamics and specification.
We generated a concrete MDP for this example in which the energy consumption depends only on the state and there are multiple local minima.
A post-posed shield was synthesized from $\spec^s_1 \wedge  \spec^s_2$, in less than a second.

\begin{figure}[!htb]
\vspace{-8pt}
\begin{minipage}{\linewidth}
  \centering
\begin{tikzpicture}[scale=.65]
\begin{axis}[
xlabel={Episodes},
ylabel={Accumulated Reward},
grid=major,
legend pos=south east,
legend style={nodes={scale=0.85, transform shape}},
]
\addplot [red,  thick] table [
x=x,
y=no_shield,
] {data/watertank.dat};
\addlegendentry{No shielding - Q}
\addplot [gray] table [
x=x,
y=no_shield_sarsa,
] {data/watertank.dat};
\addlegendentry{No shielding - SARSA}
\addplot [green, dashed] table [
x=x,
y=shield_1,
] {data/watertank.dat};
\addlegendentry{$|rank_t|=1$ w/o penalty - Q}

\addplot [color=blue, dashed] table [
x=x,
y=shield_1_sarsa,
] {data/watertank.dat};
\addlegendentry{$|rank_t|=1$ w/o penalty - SARSA}
\end{axis}
\end{tikzpicture}
\end{minipage}
\caption{The accumulated rewards per episode for different shielding settings and learning algorithms for the water tank environment. }
\label{fig:watertank plot}
\end{figure}
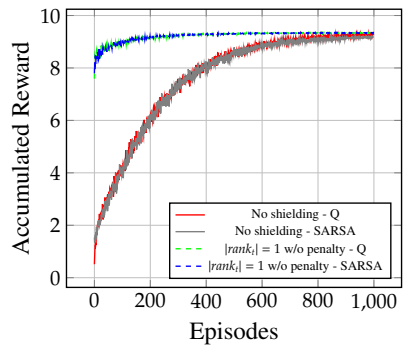

Fig.~\ref{fig:watertank plot} shows that both shielded (dashed lines) and unshielded Q-learning and SARSA experiments (solid lines) do reach an optimal policy. However, the shielded implementations reach the optimal policy in a significantly shorter time than the unshielded implementations.

\section{Conclusion}

We developed a method for reinforcement learning under safety constraints expressed as temporal logic specifications. The method is based on shielding the decisions of the underlying learning algorithm from violating the specification.
We proposed an algorithm for the automated synthesis of shields for given temporal logic specifications. Even though the
	inner working of a learning algorithm is often complex, the safety criteria may still be enforced by possibly simple means. Shielding exploits this possibility.
	
A shield
depends only on the monitored input-output behavior, the environment abstraction, and the correctness specifications -- it is independent of
the intricate details of the underlying learning algorithm.

We demonstrated the use of shielded learning on several reinforcement learning scenarios. In all of them, the shielded agents perform at least as well as the unshielded ones. In most cases, our approach even improved the learning performance.

The main downside of our approach is that in order to prevent the learner from making unsafe actions, some approximate model of when which action is unsafe needs to be available. We argue that this is unavoidable if the allowed actions depend on the state of the environment, as otherwise there is no way to know which actions are allowed. Our experiments show, however, that in applications in which safe learning is needed, the effort to construct an abstraction is well-spent, as our approach not only makes learning safe, but also shows great promise of improving learning performance.

\section*{Acknowledgements}
The authors want to thank Francisco Palau-Romero for help with some of the experiments. 
The first and last authors were supported partly by the grants AFRL  FA8650-15-C-2546, AFRL  8650-16-C-2610, DARPA W911NF-16-1-0001 and ARO W911NF-15-1-0592. The last two authors were supported partly by NSF 1617639.
The second and the fourth authors were supported by the Austrian Science Fund (FWF) through the projects RiSE (S11406-N23) and LogiCS (W1255-N23).
The third author was supported by the Institutional Strategy of the University of Bremen, funded by the German Excellence Initiative.

\bibliographystyle{splncs03}
\bibliography{ref}

\end{document}